\documentclass[aps,pra,10pt,superscriptaddress,twocolumn,nofootinbib]{revtex4-2}
\usepackage{graphicx}
\usepackage{amsmath,amsthm,amssymb,dsfont}
\usepackage{mdframed}
\usepackage{orcidlink}

\usepackage{hyperref}
\begin{document}

\title{Taxonomy for Physics Beyond Quantum Mechanics}

\author{Emily Adlam}
\affiliation{The Rotman Institute of Philosophy, Western University, 1151 Richmond Street, London, Ontario, N6A 5B7, Canada}

\author{Jonte R. Hance\,\orcidlink{0000-0001-8587-7618}}
\email{jonte.hance@newcastle.ac.uk}
\affiliation{School of Computing, Newcastle University, 1 Science Square, Newcastle upon Tyne, NE4 5TG, UK}
\affiliation{Quantum Engineering Technology Labs, Department of Electrical and Electronic Engineering, University of Bristol, Woodland Road, Bristol, BS8 1US, UK}

\author{Sabine Hossenfelder}
\affiliation{Munich Center for Mathematical Philosophy, Ludwig-Maximilians-Universit\"at M\"unchen, Geschwister-Scholl-Platz 1, D-80539 Munich, Germany}

\author{Tim N. Palmer}
\affiliation{Department of Physics, University of Oxford, UK}

\begin{abstract}
 We propose terminology to classify interpretations of quantum mechanics and models that modify or complete quantum mechanics. Our focus is on models which have previously been referred to as superdeterministic (strong or weak), retrocausal (with or without signalling, dynamical or non-dynamical), future-input-dependent, atemporal and all-at-once, not always with the same meaning or context. Sometimes these models are assumed to be deterministic, sometimes not, the word deterministic has been given different meanings, and different notions of causality have been used when classifying them. This has created much confusion in the literature, and we hope that the terms proposed here will help to clarify the nomenclature. The general model framework that we will propose may also be useful to classify other interpretations and modifications of quantum mechanics.
 This document grew out of the discussions at the 2022 Bonn Workshop on Superdeterminism and Retrocausality.
\end{abstract}
\maketitle

\section{Introduction}

 Quantum mechanics, despite its experimental success, has remained unsatisfactory for a variety of reasons, {notably due to its tension with locality, and due to the measurement problem \mbox{\cite{hance2022measprob,Adlam2023MeasProb}}}. Different authors have formulated their unease with quantum mechanics in different ways. Analysing the origin of this unease is not the purpose of this present article. The purpose is instead to sort out the confusion in the terminology used to describe this unease.

 We will below introduce a framework to distinguish between interpretations of quantum mechanics and modifications thereof. Our hope is that it may offer researchers a guide to classify their own approach. Our aim here is not to judge the promise or correctness of any of these approaches, but to make it easier to communicate among each other about what we are working on in the first place. An accompanying paper \cite{hossenfelderTK} will discuss some applications of this terminology to showcase its use.

 This paper is organised as follows. In the next section we will outline the general model framework and most importantly introduce different types of models. We believe that this classification of what we even mean by a model in and by itself will serve to alleviate the confusion of nomenclature. We will then, in Section \ref{sec:Prop} and \ref{sec:Specific} introduce some properties that such models typically have. In Section \ref{sec:Class}, we will briefly explain how to use this classification scheme, and then we conclude in Section \ref{sec:Sum}. A list of acronyms can be found in the appendix. 

\section{General Model Framework}
\label{sec:Models}

 We will start with explaining the general framework that we want to use in the following, so that we can distinguish between different types of models and their properties.

\subsection{Calculational Models}
\label{ssec:CModels}

 At its most rudimentary level, the task of physics is to provide a useful method for calculating predictions (or postdictions) for observables in certain scenarios. {(Other potential tasks associated with physics (e.g., explaining phenomena, telling us what exist, examining what is built up from what, how it changes over time and why, or producing knowledge of physical reality), are both more debatable than this, and require such a useful method themselves as a basis, so we stick with this rudimentary description as a minimal example of what the task of physics is).} A scenario is most often an experiment, and this is the situation we are typically concerned with in quantum foundations. But in some areas of physics---such as cosmology and astrophysics---one does not actually conduct experiments, but instead {one observes} a natural variety of instances that occurred in the past. For this reason, we will use the more neutral term ``scenario.'' A scenario is not itself a mathematical structure; it is the real-world system that we want to describe using mathematical structures.

\begin{figure}
    \centering
\includegraphics[width=\linewidth]{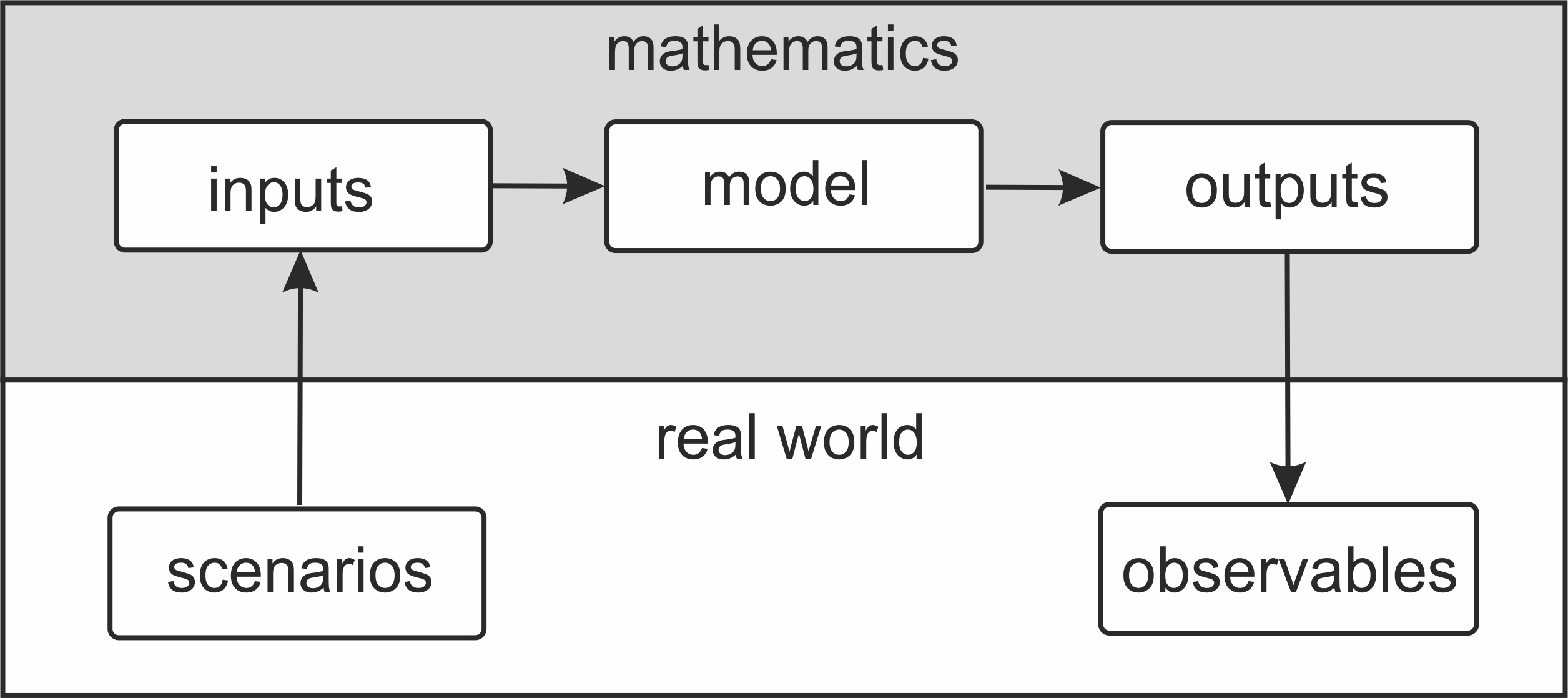}
    \caption{The logical flow of the model framework.}
    \label{fig:Models}
\end{figure}

 The basic logical flow of such a calculation is outlined in Fig.~\ref{fig:Models}. We will call the center piece of this calculation a {\bf calculational model}, or {\bf c-model} for short. We do not call it a ``computational model'' because ``computational'' suggests a numerical simulation, which is an impression we want to avoid. A calculational model could be numerical, but it could also be purely analytic. For more on the relation between calculational models and computer models, see Section \ref{ssec:comp}.
 
 We will in the following use a notation in which curly brackets $\{\cdot \}$ denote sets of mathematical assumptions. The set $\overline{\{\cdot \}}$ is the set of all assumptions that can be derived from $\{\cdot \}$ (including the elements of $\{\cdot \}$ themselves).

 The different components of the modelling framework have the following properties:

\begin{quote} 
 {\bf Inputs (of a calculational model):} The inputs ${\cal I}$ of a calculational model are an (at most countably infinite) set of mathematical assumptions---each of which is an input, $I_i$---that describes the scenario. To be part of the inputs, an assumption must differ between at least two scenarios. We will denote this set as {${\cal I}:=\{I_i ~|~ i \in K \subseteq \mathbb{N}^+ \}$, were $K$ is the (at most countably infinite) index range of the assumptions.}
\end{quote}

{To aid readability, and because the index set will not matter in the following, we will from here on assume that all sets are at most countably infinite and suppress the index range. That is, we will just write $\{I_i \}$ rather than$\{I_i ~|~ i \in K \subseteq \mathbb{N}^+ \}$}.

 Loosely speaking, you can think of the inputs as the mathematical representation of a scenario. A typical example might be the temperature in the laboratory, or the frequency of a laser. But the inputs of a model do not necessarily have to correspond to definite observable properties of the scenario. They could also be expressing ignorance about certain properties of the scenario and thus be random variables with probability distributions, or they might indeed be entirely unobservable. We will come back to this point in Section \ref{ssec:atemporal}.

The inputs of a c-model are often assignments of values to variables. For example, a c-model may work with an unspecified variable $x$ that is an element of some space. The input may then assign the value $x=3$ to this variable for a specific scenario. For such an input, we will refer to the value it assigns as the {\bf input value}. We want to stress that the input value is not the same as the input. The input is ``$x=3$'', the input value is ``$3$''.

A typical example for a c-model that we are all familiar with would be the Harmonic oscillator. In this case, the model would be the differential equation $m \ddot x = -k x$ and a complete set of inputs would be value assignments for $k$ and $m$, plus two initial values for, say, $x(t=0)$ and $\dot{x}(t=0)$.

 But inputs of a calculational model are not necessarily value assignments. They might also be constraint-equations, or boundary conditions, or something else entirely. For example, when studying stellar equilibrium, it is quite common to enter an equation of state as the input to the Tolmann-Oppenheimer Volkov ({\sc{TOV}}) equations. In this case, the {\sc{TOV}} equations are the same for all stellar objects that one may want to consider; they are hence part of the model. The equation of state, on the other hand, changes from one type of star to another; it is hence part of the input.
 
 While our main interest is in models that describe the real world, it is also possible to study a model's properties with scenarios that do not exist in reality. We will refer to those as {\bf hypothetical scenarios}. They include, but are not limited to, counterfactual realities, as well as universes with different constants of nature. {(Note Frigg and Nguyen \mbox{\cite{frigg2020modelling}}, amongst others, have also discussed representation in scientific modelling.)}

\begin{quote} 
 {\bf Calculational model:} A set ${\cal C}$ of mathematical assumptions $A_x$ that are independent of the scenario. We will denote this set as $ {\cal C}:=\{A_k \} $.
\end{quote}

\begin{quote} 
 {\bf Setup:} The union of a calculational model and its inputs. ${\cal F}:= {\cal I} ~\cup~ {\cal C}$.
\end{quote}

\begin{quote}
 {\bf Outputs (of a calculational model):} All mathematical statements that can be deduced from setup, but from neither the model nor its inputs in isolation $O= {\overline F} \setminus (\overline{\cal I} ~\cup~ \overline{\cal C}) $.
\end{quote}

 Predictions from observables are obtained from the outputs of the model. However, not all outputs of a model need to be observable. The prime example is quantum mechanics, where the outputs contains the time-evolution of the wavefunction, but the wavefunction itself is not observable. But there are many other examples, such as the production of gravitational waves by a black hole merger. Given suitable inputs, the model (General Relativity) will output a mathematical description for the creation and propagation of gravitational waves, but we only measure their arrival on Earth, and only measure that through the waves' influence on matter, which is a small part of all the model outputs. And like the inputs, the outputs of a model do not necessarily have to result in definite values for observables; they could instead merely give rise to probability distributions for values of observables.

 For a calculational model to be useful, we further need a prescription to encode a scenario with specific inputs, and a way to identify the observables from the outputs. This identification, since it is not restricted to the mathematical world, is not one that we can strictly speaking define. Science, after all, is not just mathematics. What property this identification with the real world must have is an interesting question, but it is one of many that we will not address here, because it is not relevant for what follows.
 
 When using hypothetical scenarios, these have to be distinguished from the setup as a matter of definition. If the hypothetical scenarios are not identified as such by definition, it becomes impossible to tell whether one changes the hypothetical scenario or the model.\footnote{An example for this is the case of the Standard Model with its fundamental constants not fixed but taken as variable inputs. This cannot be a correct model for scenarios in our universe because in our universe the constants are constant, hence they can't be inputs. One can, however, consider hypothetical scenarios with different values of the constants, often interpreted as a type of multiverse. But of course one could alternatively interpret these hypothetical scenarios as different versions of the Standard Model that, alas, happen to not agree with our observations. The point is that whether the Standard Model with fundamental constants that don't agree with our observations describes a hypothetical alternative universe, or is just a wrong model for our universe, is a matter of definition.}

 Another property, relevant for our purposes, that we want the setup of a calculational model to have is: That they are {\bf irreducible}, in the sense that we cannot split the combined set ${\cal C}:=\{I_j, A_k\}$ into two sets ${\cal C}_1:= \{I^1_{j^1}, A^1_{k^1}\}$ and ${\cal C}_2:=\{I^2_{j^2}, A^2_{k^2}\}$, where $\{I_{j} \} = \{I^1_{j^1} \} ~\cup~ \{I^2_{j^1} \}$ and $\{A_{k} \} = \{A^1_{k^1} \} ~\cup~ \{A^2_{k^2} \}$, so that both ${\cal C}_1$ and ${\cal C}_2$ are each setups of calculational models and the combination of their outputs is the same as the outputs of ${\cal C}$. A simpler way to put this is that, if we split the setup of an irreducible model into two, some of the output can no longer be calculated. This approach is reminiscent of identifying particles in quantum field theory from the irreducible representations of the Poincar\'e group.

 We need this requirement because otherwise we could just join different setups to form a new one, which would make it impossible to classify any. Note that this does not mean the composition of two setups is not a setup. On the contrary, a composition of two setups {\emph{is}} a setup, it's just that the combined setup is no longer irreducible. The issue is, if we allowed reducible combinations of setups, then we could not meaningfully assign properties to them. It would be like asking which fruit is fruit salad. Once we have however succeeded in identifying properties of irreducible setups, we can join those with the same properties together, and meaningfully assign the same property to the reducible setup. Using the above fruit example, if we have identified several fruits as apples, we can join them and be confident we have apple salad.

 A particularly relevant case of a reducible setup is one in which some inputs or assumptions can be removed without changing anything about the outputs. This may be because the assumptions are not independent (in the sense that some can be derived from the others), or because an assumption is simply not used to calculate the outputs. 

 One might be tempted to add to the requirements of a model that its assumptions are consistent {(given problems such as the Principle of Explosion with inconsistent models \mbox{\cite{macfarlane2020philosophical}})}. However, as is well known, for recursively enumerable sets, G\"odel's theorem {\cite{godel1931formal}} tells us that we cannot in general prove the consistency of the assumptions. We might then try to settle on the somewhat weaker requirement that at least the assumptions should not be known to be inconsistent. However, it sometimes happens in physics that a model works well in certain parameter ranges despite being inconsistent in general. An example may be the Standard Model without the Higgs field and its boson {\cite{smith1973high}. For this reason, we will here not add any requirement about consistency. One may justify this by taking a purely instrumental approach. We only care whether a set of assumptions is any good at describing observations.

 The setup of a calculational model whose inputs are all value assignments that, alas, have not been assigned is what is usually called a model in the causal models literature \cite{pearl2009causality,glymour2001causation}.

\subsection{Mathematical Models}

 We defined a calculational model and its inputs as sets of mathematical assumptions. Such sets can be expressed in many different ways that are mathematically equivalent. 
 We will call an equivalence class of calculational models a {\bf mathematical model, or m-model} for short, that is, the term m-class means the same as m-model. We will denote this equivalence class with $[\cdot]_{\rm m}$ and refer to it as an {\bf m-class}; it is a set of sets. For example, if ${\cal C} = \{A_k\}$ is a calculational model, then ${\cal M} := [{\cal C}]_{\rm m}$ is the m-model that encompasses all mathematically equivalent formulations
 
\begin{equation}
 [{\cal C}]_{\rm m} := \{ \{B_j\}: \{B_j\} \Leftrightarrow \{A_i \} \}~.
\end{equation}

 Calculational models in the same m-class will be called {\bf m-equivalent}. By a mathematical equivalence of the sets (``$\Leftrightarrow$''), we here mean that either one can be derived from the other. We thereby adopt the notion of model equivalence advocated by Glymour \cite{glymour1970theoretical,glymour1981theory}}\footnote{Glymour uses the term ``theory equivalence'', not ``model equivalence''. This is not a relevant distinction for the purposes of this present paper: please see section \ref{ssec:Theories} for discussion.} 

 It was argued by Weatherall \cite{weatherall2019part1,weatherall2019part2} that mathematical equivalence might over-distinguish models in certain circumstances, and that it might be better to use a weaker form of structural equivalence based on category theory.
 We do not use this proposal of categorial equivalence here, not because we object to it, but because it is neither widely accepted nor understood. 
 Most importantly, it is at present not practical because few physicists would know how to apply it. Mathematical equivalence, in contrast, is widely understood and comparably straightforward (though not necessarily easy) to prove.

 Among the models we will discuss here, mathematical models are closest to what physicists typically mean by a ``model''. They do not distinguish between the particular mathematical formulations that one might use to make a calculation. For example, we could express Maxwell's Equations using differential forms with ${\rm d} \star, {\rm d} \wedge$ and $\star {\rm d}$ operations, or we could do it with the more old-fashioned $\vec \nabla, \vec \nabla \cdot$ and $\vec \nabla \times$ operators. These two definitions would be two different calculational models, but the same mathematical model. A similar equivalence covers switching from the Schr\"odinger picture to the Heisenberg picture in quantum mechanics. We believe that most physicists would not call these different models.

 The inputs of a mathematical model are likewise an equivalence class: It is the class of all sets of assumptions which change with the scenario that, together with the mathematical model, produce equivalent outputs. 

 The inputs of a mathematical model are usually not just the set of assumptions that are mathematically equivalent to to the inputs of one of the calculational models. This is because the assumptions of the model itself may render certain inputs equivalent that are not equivalent without the model. An example that will be relevant for what follows is that a model with a time-reversible evolution operator will produce identical outputs for input states at times $t_1$ and $t_2 > t_1$, if the input state at $t_2$ is the forward evolution of the state from $t_1$. In this case the input is m-equivalent, though the inputs at different times are not equivalent to each other without the model. 

 We will refer to different calculational models as {\bf representations} of their m-class.
\begin{quote}
 {\bf Representation (of an m-model):} A calculational model {within the m-class of that m-model}. 
\end{quote}

 In \cite{argaman2018lenient}, {a similar concept, that of two c-models within the same m-class, were referred to as ``reformulations'' of each other}.

\subsection{Physical Models}

 The previous definitions do the heaviest lifting for the model framework, and the remaining ones are now straight-forward. It can happen that mathematical models are different, but they nevertheless give rise to the same observables for all scenarios. We will combine all such mathematical models into yet another, even bigger class and say they constitute the same {\bf physical model, or p-model} for short. We will denote this equivalence class with $[\cdot]_{\rm p}$ and refer to it as the {\bf p-class}. 
 Note that we could take either the p-equivalence class of a computational model or that of a mathematical model. Models in the same p-class will be called {\bf p-equivalent}. 
 
 By calling these models physical, we do not mean to imply that we only consider observable properties as physically real. The reader should not take our nomenclature to imply any statement about realism or empiricism. We call them physical just because they describe what physicists in practice can distinguish with physical measurements.
 
 This now gives us a way to define what we mean by an interpretation:
\begin{quote}
    {\bf Interpretation (of a p-model):} {Any one of the} mathematical models in the p-class. 
\end{quote}

Note that according to this nomenclature\footnote{{(Note others have given different definitions of interpretations for quantum mechanics, e.g., \mbox{\cite{Muller2013Circum}}.)}}, each interpretation has its own representations. That is, a c-model is a representation of an m-class, but more specifically it is a representation of a particular interpretation.

 For example, if we take the Copenhagen Interpretation (CI) in one of its common axiomatic formulations {(e.g., as given in \mbox{\cite{zurek2003decoherence}})}, that set of axioms will constitute a particular representation, hence, a c-model. The Copenhagen Interpretation per se is the class of all mathematically equivalent models. We can then further construct the physical equivalence class of the Copenhagen Interpretation, which we will in the following refer to as Standard Quantum Mechanics ({\sc SQM}:=[CI]$_{\rm p}$). Any other mathematical model that is physically but not mathematically equivalent to the Copenhagen Interpretation is also an interpretation of {\sc SQM}.

For the purposes of this paper we will not need to specify exactly what we mean by Copenhagen Interpretation. However, we will take it to be only a first-quantised theory. That is, it is not a quantum field theory. If one were to specify further details, one would in particular have to decide whether one considers the relativistic version, or only the non-relativistic case, as some alternative interpretations work only non-relativistically\cite{pittphilsci20537}.

 That we have interpretations and not just representations is a possibility which appears in quantum mechanics because it has outputs that are unobservable in principle. A different computational model which affects only unobservable outputs might not be mathematically equivalent and yet give rise to the same physics.

\subsection{Empirical Models}\label{ssec:EmpModels}

\begin{figure}
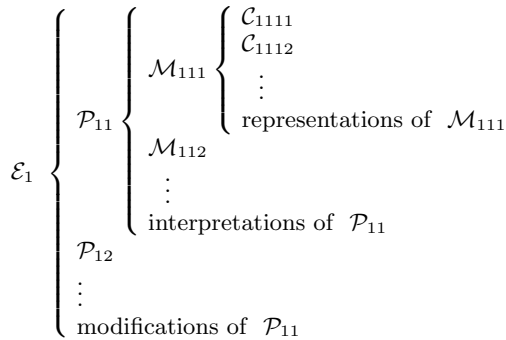

\begin{eqnarray*} {\cal E}_1~
\left\{ \begin{array}{l} 
{\cal P}_{11} 
\left\{ 
\begin{array}{l} 
{\cal M}_{111} 
\left\{ \begin{array}{l} 
{\cal C}_{1111} \\
{\cal C}_{1112}
\\~~ \vdots \\
\mbox{{representations of}}~~ {\cal M}_{111} 
\end{array} \right.
\\ 
{\cal M}_{112}\\~~\vdots\\
\mbox{{interpretations of}}~~ {\cal P}_{11} 
\end{array} \right.
\\
{\cal P}_{12}\\
 \vdots\\
\mbox{{modifications of}}~~ {\cal P}_{11} 
\end{array}
 \right.
\end{eqnarray*}
 \caption{A mathematical model ${\cal M}_{ijk}$ is an equivalence class of computational models ${\cal C}_{ijkl}$. A physical model ${\cal P}_{ij}$ is an equivalence class of mathematical models ${\cal M}_{ijk}$. An empirical model ${\cal E}_i$ is an equivalence class of physical models ${\cal P}_{ij}$. \label{fig:Tree}}
\end{figure}

 Finally, we define an {\bf empirical model, or e-model}, ${\cal E}$, as the class of all physical models that cannot be distinguished by observations so far. We will denote this class as $[\cdot]_{\rm e}$, refer to it as an {\bf empirical class or e-class}. Models in the same e-class are {\bf empirically equivalent or e-equivalent}. We will call anything in the same empirical class as {\sc SQM}, but not in the same physical class, a modification of quantum mechanics:

\begin{quote} 
 {\bf Modification or Completion or Extension (of a p-model):} Another p-model that is not physically equivalent but so far empirically equivalent.
\end{quote}
 Fig.~\ref{fig:Tree} summarises the relations between these various types of models. 

 We here use the terms modification, completion, and extension loosely, and will not distinguish them below because it is not necessary for what follows. In the literature these terms are used with a somewhat different meaning. A modification or extension of a model typically employs similar mathematical structures, whereas a completion introduces new mathematical structures. However, the distinction between the two is not always clear. Again, just how to distinguish them is an interesting question in its own right, but not one we will address here. 

With a slight abuse of terminology that is however unlikely to cause misunderstandings, we can also use the above definition to refer to an m-model or a c-model as a modification. That is, an m-model (c-model) is a modification of another m-model (c-model) if the two are not physically equivalent, but so far empirically equivalent. 

 One could further refine the class of empirically equivalent models by specifying various sets of experiments. For example, we could ask what models can be distinguished by experiments in the near future, and what exactly we mean by near future. Or we could talk about models that can be distinguished by experiment in principle, and then discuss whether, say, waiting one hundred billion years is possible in principle. Or whether it is possible in principle to make measurements inside a black hole, and so on. All of these are very interesting questions; however they will not concern us in the following, so we will not elaborate on them. 

\subsection{Theories}
\label{ssec:Theories}

 We will in the following not distinguish between models and theories. Loosely speaking, a theory is a class of models that can be used for a large number of scenarios, congruent with the semantic approach of Suppes \cite{Suppes1961Models} and van Fraassen \cite{VanFraassen1989laws}. However, physicists don't tend to use the terms theory and model in a well-defined way. 

 For example, we use the terminology of Quantum Field Theory in general, and the Standard Model in particular. We refer to General Relativity as a theory, and to $\Lambda${\sc CDM} as a model. This agrees with what we said above. 
 But we also refer to Fermi's theory of beta decay, and the {\sc BCS} theory, though those would better be called models. To make matters worse, we also sometimes call supersymmetry a theory, when it's really a property of a class of m-models, and so on. For the purposes of this paper, we will not need to distinguish theories from models, so we will not bother with a precise definition of the term ``theory.''

\subsection{Computer Models and Simulations}
\label{ssec:comp}

 Another type of model which is commonly used by scientists of all disciplines are computations performed on computers. There are two ways to think of these models. 

 One is that they are really simulations that represent one real-world system with another real-world system. That is to say, they are not models in the sense that we have considered here---the models we considered here are mathematical. A suitably programmed computer is instead a physical stand-in for the system one wants to make a prediction for. This is also how quantum simulations work \cite{georgescu2014quantum,hangleiter2022qsim}. 

 However, there is another way to look at computational models, which is to use their program as a definition for a calculational model. This then falls into the classification scheme we discussed above. But in this case, the computational model is typically not mathematically equivalent to the analytical, calculational model that one used for a scenario. This is because computers are in almost all cases digital, and only approximate the continuum. The exception are certain types of analog computers, which however can better be understood as simulations again. 

 That is to say, when one wants to classify a computer model according to our scheme, one should take its algorithmic definition as a set of assumptions, and use that to define a calculational model and its inputs. This calculational model, defined by its executable algorithm, will be different from the calculational model that uses an analytic expression. 

 (One can then further ask to what accuracy will the algorithm, when executed on a physical computer, approximate the output of the analytical model. This is a relevant point which was previously brought up in {\cite{wharton2020colloquium}}. While an analytically defined calculational model might be time-reversible, an algorithmic approximation of it run on a computer will in general no longer be time-reversible. This is because errors will build up differently depending along which direction of time one runs the algorithm. This is particularly obvious for time-evolutions which are chaotic. In such cases, the forward-evolution and the backward-evolution of the algorithm as executed on the computer will actually be two different calculational models, and both are different from the analytical calculational model that they approximate).

\section{General Model Properties}
\label{sec:Prop}

 In this section, we will discuss some properties that we can assign to models and their inputs. The point of this section is to specify which properties are m-class properties (do not change with mathematical redefinitions), and which are c-class properties (can change with mathematical redefinition). 
 
 We want to stress that we will {\emph{not}} prove that these assignments are correct. To do that, we would have to add many more assumptions (e.g., about what we mean by space, and time, and measurements, and so on). What we will do instead is specify what we believe is the way that an expression has been previously dominantly used in the literature, and this specification will then in turn imply properties for the concepts we did not specify.

 To make this procedure clear, if we state for example that the property of "time-reversibility" does not change with a mathematical redefinition, then this implicitly means: If it did change, we would not refer to the property as time-reversibility. That is, there are certain properties of models which we want to not depend on just exactly how we write down the mathematics.

 Quite possibly, some readers will disagree with our some of our assessments. This is fine. Our aim here is not to end the discussion, but to propose a language in which we can talk about these properties in the first place.

 The term ``model'' without further specification (c/m/p/e) will refer to any type of model. 

\subsection{Atemporal Properties}
\label{ssec:atemporal}

 Following the terminology of Cavalcanti and Wiseman \cite{cavalcanti2012bell}, we will call an m-model ``deterministic'' if its observables can be calculated with certainty from the outputs obtained from the setup:
\begin{quote} 
    {\bf Deterministic:} An m-model is deterministic {iff}\footnote{{We use the common mathematical abbreviation ``iff'' for ``if and only if''.}} the probabilities for predictions of observables are all either 0 or 1.
\end{quote}
 This property was termed ``holistic determinism'' in \cite{adlam2021determinism} and differs from a more common definition of determinism, that connects one moment in time with a later one. We will elaborate on this in the next subsection. For now, we further follow \cite{cavalcanti2012bell} and distinguish determinism from predictability:
\begin{quote} 
    {\bf Predictable:} A m-model is predictable {iff} it is deterministic, and the inputs are all derived from observable properties of the scenario.
\end{quote}
 Both determinism and predictability are m-model properties, because we expect a redefinition of the mathematics that one uses to process inputs to not change predictions for observable output. If that was the case, we would assume something is wrong with our idea of what is observable.

 The distinction between deterministic and predictable is that a model may have inputs that are unknowable in principle. Typically these are value assignments for variables that are usually referred to as ``hidden variables''. A model with such hidden variables, according to the above terminology, may be deterministic and yet unpredictable.
\begin{quote}
 {\bf Hidden Variables:} Hidden variables, that we will denote $\kappa$, are input values to a c-model that cannot be inferred from any observation on the scenario. 
\end{quote}

 We want to stress that these hidden variables are in general not localised and sometimes not even localisable in any meaningful way. We will say more about localisable variables in the next subsection, but a simple example is that we could use Fourier-transforms of space-time variables, or just extensive quantities that are properties of volumes. Note that hidden variables are defined for c-models, not for m-models. This is because hidden variables can be redefined into ({not necessarily local}) random variables with an equivalent mathematical outcome. That is to say, mathematically it makes no difference whether a variable was unknowable or indeed random. 

 While it may sound confusing at first to distinguish determinism from predictability, it will be useful in what follows. Indeed, the reader may have recognised that Bohmian Mechanics is deterministic yet not predictable. In Bohmian Mechanics, observables can be calculated with certainty if the inputs are specified, yet the inputs are also assumed to be partly unobservable in principle, so predictions can still not be made with certainty. Since determinism is a property of an m-model that cannot be removed with a re-definition, it follows that Bohmian Mechanics is not a representation of {\sc SQM}. We elaborate on this further in the companion paper \cite{hossenfelderTK}.

\subsection{Local and Temporal Properties}
\label{ssec:LocalTemporal}

 We will now add some properties of models that we commonly use in physics, starting with those that refer to locations in space and time. Clearly we can only speak of locations in space and time if a model has some notion of space and time, and a distance measure on them, to begin with. In the simplest case, that would be the usual space-time manifold and its metric distance. But it could alternatively be some other structure that performs a similar function, such as a lattice, or a discrete network with a suitably defined metric. 

 To make sense of space and time we will in the following consider only a restricted m-class associated with a calculational model, that is one in which space and time are consistently defined throughout the entire class. To see why this is necessary, imagine if we were to redefine one direction of ``space" as ``time" and vice versa. This is mathematically possible, but it makes no physical sense. It would screw up any notion of locality and causality just by nomenclature. We will hence assume that space and time and a distance measure on them are consistently defined throughout the m-class.

 However, sometimes a model just does not have a description of space-time or a notion of locality. This might sound odd, but is not all-that-uncommon in the foundations of quantum mechanics, where many examples deal with qubit states that do not propagate and are not assigned space-time locations \cite{Spekkens2007ToyModel}. We will refer to such models as alocal:

\begin{quote}
 {\bf Alocal:} An m-model that does not have a well-defined notion of space, space-time, locality, or propagation. 
\end{quote}

 One cannot fix a lack of definition by switching to a different but mathematically equivalent definition, so if a c-model is alocal, then so will be any other c-model its m-class. 

 For alocal models, we cannot make any statements about whether they are local or not. Similarly, a model may just not have a notion of time or a time-evolution. This again is not all that uncommon in the foundations of quantum mechanics. Many elaborations on correlations and classicality bounds, for example, do not specify a time-evolution for states; and the process matrix formalism is specifically designed to study possible quantum processes without a predefined time order \cite{articleoreshkov,Oreshkov2}. 
\begin{quote}
 {\bf Atemporal:} An m-model that does not have a well-defined notion of time.
\end{quote}

 For models which have a proper notion of space-time, and a distance measure on it, we then want to identify how local they are. For this, we first have to identify the input values that can be assigned to space-times which Bell coined ``local beables'' \cite{bell1975theory}. According to Bell, local beables ``can be assigned to some bounded space-time region''. We will take ``assigned to'' to mean that the variable is the value of a function whose domain is space-time, or whatever the stand-in for space-time is in the model at hand. {Note ``region'' might be a point set.}

 For example, if space-time is parameterised by coordinates $(\vec x,t)$, and we have a function $f: (\vec x, t) \to \mathbb{C}$, then $f(\vec x, t)$ is a local beable. The domain of the {\sc SQM} wavefunction is generically configuration space and it is therefore not a local beable, though under certain circumstances local beables can be obtained from it (such as the single-particle wavefunction in position space). Local beables are not necessarily observable.

 Bell's definition of a local beable makes the assignment to a space-time region optional (``can be assigned''). However, if the assignment was optional, then it could be omitted, and a model with an assumption that can be omitted is reducible. Since we only deal with irreducible models, we therefore already know that if a variable is assigned to a space-time region, then this assignment is actually necessary. For this reason we define
 
\begin{quote}
 {\bf Local Beable:} An input value of a c-model that is assigned to a compact region of space-time.
\end{quote}
 In the following, ${\cal I}(A)$ will refer to the local beables in space-time region $A$.

 It becomes clear here why we required our models to be irreducible: so that it is actually necessary, and not just possible, to assign the local beables to space-time regions. It has for example been rediscovered a few times independently \cite{brassard2019parallel,ciepielewski2020superdeterministic} that any theory can be made local (in pretty much any reasonable definition of the term) by copying the local beables from any space-time location into any other space-time location and postulating them to ``be there'' without any other consequences. 

 Concretely, suppose we have localised variables $f(x,t)$ over a space-time manifold $M$, then we can just define the state $f(x,t) \otimes f(x',t') \in M \otimes M$ and call that a local beable at $(x,t)$, just that now we have the information about any $(x',t')$ also available at the same point. Depending on perspective, one might consider such a model as either ultra-local or ultra-nonlocal. It is ultra-local in that we do not need to know the state at any other location than $(x,t)$ to calculate the time-evolution. It is ultra-nonlocal because any point in space-time contains a copy of the entire universe.

 A similar problem would occur with parameters, such as $\hbar$, that could be transformed into fields $\hbar: (x,t) \to \delta(x-x')\delta(t-t')$ and then be used as inputs with random locations, leading to seemingly `non-local' interactions in the Hamiltonian.

 Such procedures to localise beables create new c-models that are in different m-classes, because the copying procedure is an additional assumption that could not have been derived from the original version of the c-model. Since such a localising assumption is unnecessary to calculate outputs, the setup of such a model is reducible and, as we have previously remarked, properties of reducible setups are ambiguous. 

A word of caution is in order here. As mentioned above in the elaboration on alocality, many discussions of violations of Bell's inequalities do not explicitly state locality assumptions. These assumptions might hence appear unnecessary. This is indeed correct if one merely looks at the inequality. However, if one wants to draw conclusions about locality from a violation of Bell's inequality, one needs to at least make a statement about which parts of the experiment are causally connected or space-like separated. That is, we suspect that many alocal models do implicitly require locality statements to arrive at the desired conclusions. One should not be deceived by only looking at the explicitly stated assumptions.

 Having introduced local beables, we can now define an alternative:
\begin{quote}
    {\bf All-At-Once (AAO) input:} Input of a c-model that constrains or defines properties of {at least one local beable} for at least two temporal regions, {and that cannot be decomposed into inputs in separate temporal regions}.
\end{quote}
 Typical examples of such inputs are event relations, consistency requirements for histories, evolution laws, temporal boundary conditions, or superselection rules. Something as mundane as an integral over time that depends on the scenario would also be an all-at-once input. 
\begin{quote}
    {\bf All-At-Once (AAO) model:} A c-model that uses all-at-once input.
\end{quote}
 The use of all-at-once input is \textit{a priori} a property of c-models. That is to say, it might be possible to reformulate a model with {\sc AAO} input into a mathematically equivalent model that does not have this property.
 
 The principle of least action in classical mechanics, for example, uses All-At-Once input (the action), but given that the Lagrangian fulfills suitable conditions, the principle can equivalently be expressed by the Euler-Lagrange equations which do not require AAO input. Another example may be the cosmological model introduced in {\cite{carroll2017nonlocal}}, which uses a constraint on the space-time average of the Lagrangian density. The authors refer to this as non-local, and in some sense it is, but it is more importantly also an all-at-once input.

\begin{figure}[ht]
    \centering
    \includegraphics[width=\linewidth]{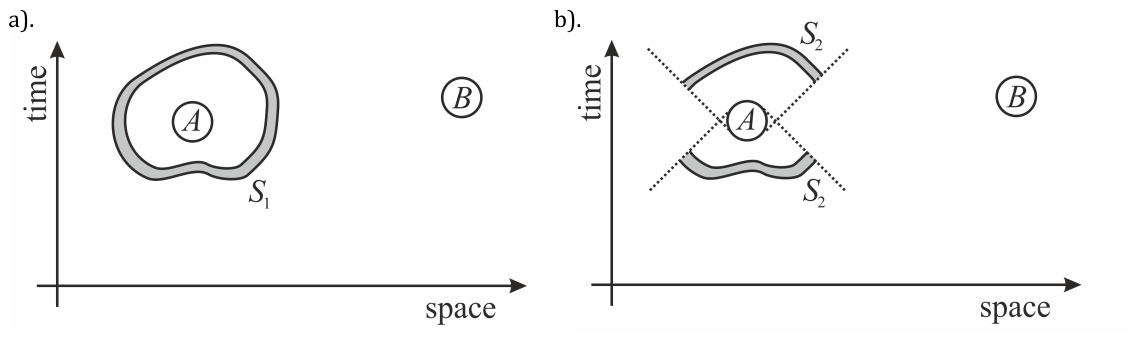}
      \vspace*{-0.5cm}
    \caption{a). Continuity of Action. b). Strong Continuity of Action.}
    \label{fig:CoA012}
\end{figure}

 Now that we have localised variables, we can define a notion of locality. We will first leave aside causality and introduce a notion of Continuity of Action, as done in \cite{wharton2020colloquium}. The idea is that if one wants to calculate the outputs ${\cal O}(A)$ for a region $A$, then it is sufficient to have all the information on a closed hypersurface, $S_1$, surrounding the region, and adding local beables from another region outside $S_1$ provides no further information. 
\begin{quote}
 {\bf Continuity of Action (CoA, locality):} An m-model has continuous action or fulfils Continuity of Action or is local {iff} it fulfils the condition $P({\cal O}(A) | {\cal I}(S_1), {\cal I}(B)) = P({\cal O}(A) | {\cal I}(S_1))$ for any region $S_1$ that encompasses $A$ but not $B$ (see Fig.~\ref{fig:CoA012}a.)
\end{quote}
\begin{quote}
 {\bf Non-locality:} An m-model is non-local {iff} it violates CoA.
\end{quote}
 Note that fulfilling CoA does not mean that the outputs in $A$ are determined by the input values on $S_1$ to begin with. It's just that adding information from $B$ doesn't make a difference. 

A natural question to ask here is just how small the regions should be for the model to have continuous action. This is a most excellent question but luckily one that we do not need to answer because it resides on the level of empirical adequacy. A model in which variables cannot be localised sufficiently much will simply not reproduce certain observations. (For example, a model with a spatial resolution so rough that it cannot distinguish two detectors from each other can only give a total probability for both, not for each separately.) The typical size of the regions is what is often referred to as the coarse-graining scale of a model. 

A subtlety of the definition for Continuity of Action was pointed out in \cite{palmer2019bell}, which is that some combinations of inputs may be mathematically possible, but are not in the physical state space of the model and hence do not correspond to any scenario which can occur in reality. This might happen for example because certain combinations of variables are just forbidden by a mathematical constraint (as happens with the Fermi exclusion principle). In this case, one might have a situation where e.g.\ $P({\cal I}(S_1), {\cal I}(B)) =1$ and $P({\cal I}(S_1), {\cal I}(B)') =0$ just because the latter combination is incompatible with the assumptions of the model \cite{Hance2022Supermeasured}. It would then seem that $P(A|{\cal I}(S_1), {\cal I}(B)) \neq P(A|{\cal I}(S_1), {\cal I}(B)')$ and CoA is violated. However, as argued in \cite{palmer2019bell}, it is rather meaningless to talk about violations of locality for configurations which do not physically exist. Hence, if a model has input constraints, one should only apply the locality requirement to inputs that lead to physically possible scenarios. 

 We define CoA as an m-model property because, if it could be removed with a mathematical redefinition, we believe most physicists would not accept the definition as meaningful.

 The term ``non-locality'' has been used to refer to many other definitions in the foundations of physics in general, and quantum mechanics in particular. For example, in field theories, non-locality often refers to dynamical terms of higher order, a definition that is also used in General Relativity {\cite{hehl2009nonlocal}}. In quantum field theories, non-locality usually refers to the failure of operators to commute outside the light cone. Even in quantum foundations, non-locality may refer to different properties. For example, entanglement itself is sometimes considered non-local despite being locally created {\cite{Croke2022Nonlocal}}. The latter in particular has created a lot of confusion, because while it has been experimentally shown that entanglement is a non-local correlation and does in fact exist, this does not imply that nature is non-local in the sense that the term has been used in Bell's theorem, which has of course not been shown \cite{Hance2022ComNatPhys}.

 Surveying all those different notions of non-locality is beyond the scope of this present work. However, we want to stress that as definitions, {none} of these notions of non-locality are wrong. They are just that: definitions. We chose our definition to resemble closely the ``spooky action at a distance'' that Einstein worried about.
 
 According to our definition, a calculational model fulfills CoA even if it can't be directly evaluated whether it fulfils the requirement on the conditional probabilities, so long as a mathematically equivalent reformulation of the model fulfills it.
 The most relevant example for our purposes is that $[{\rm CI}]_{\rm m}$ does not fulfil CoA. This is because making a measurement in $B$ provides information about the measurement outcome in $A$ that was not available in $S_1$. This is Einstein's ``spooky action at a distance''.
 
That is, with the terminology we have introduced so far, the Copenhagen Interpretation and all mathematically equivalent formulations are equally non-local. The question is then merely whether this is something that we should worry about. If one can understand the wavefunction as an epistemic state (a state of knowledge), then its non-local update is not \textit{a priori} worrisome. 
 
 Continuity of action loosely speaking means that localisable variables can only influence their nearest neighbours. However, it does not require that this influence lies within the light cones. To arrive at a stronger condition, we will therefore now as is customary assign the light-cones and their insides to a space-time region, $A$. We will denote with $L(A)$ the union of all space-time points that are light-like or time-like related to any point in $A$.

\begin{figure}[t]
    \centering
    \includegraphics[width=\linewidth]{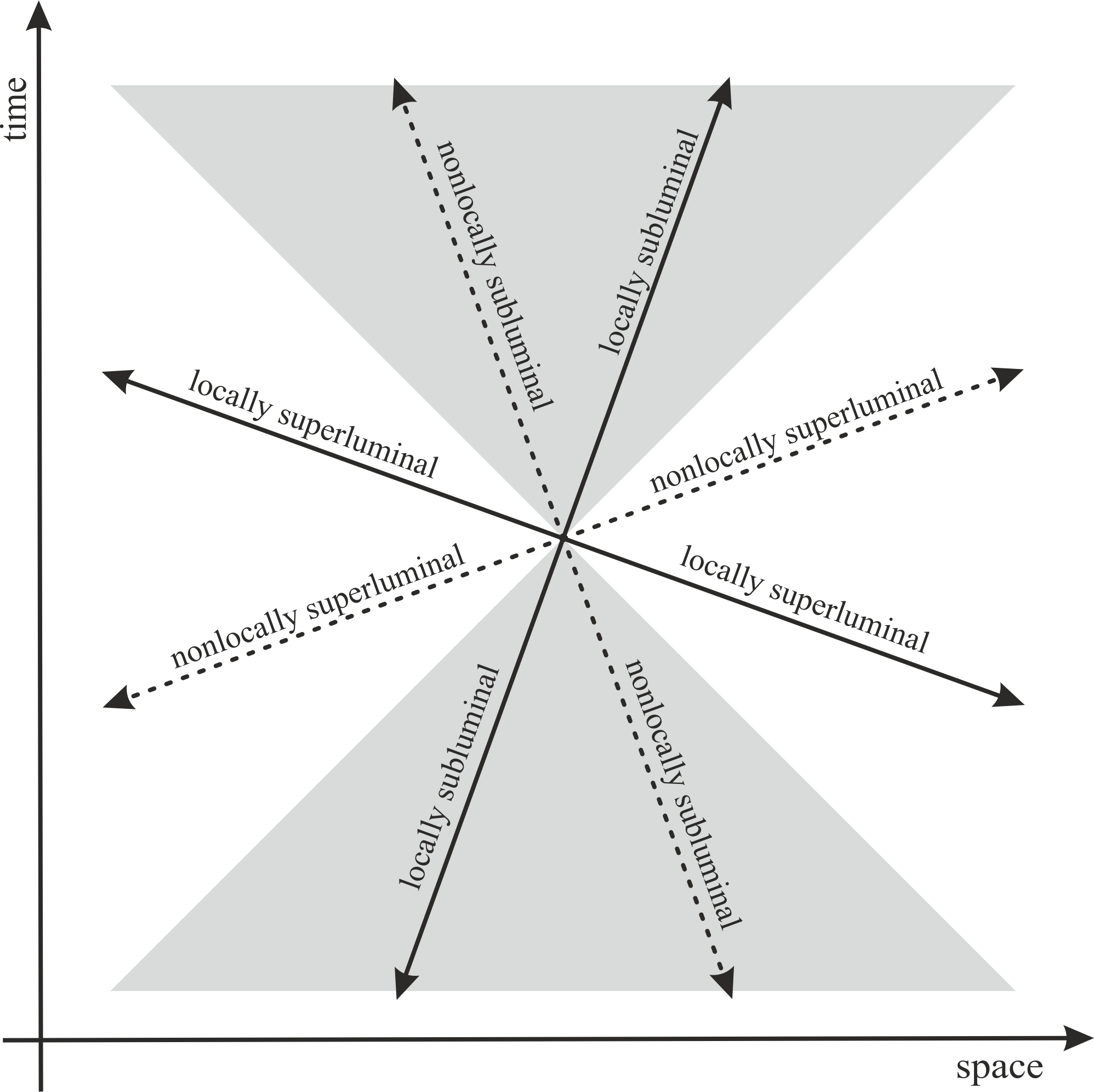}
      \vspace*{-0.2cm}
    \caption{Difference between local, non-local, and superluminal. {Note that local and nonlocal are here distinguished by having solid/dotted lines respectively, and that each arrow is only one representative for the entire quadrant (e.g., one can imagine a reflected version of the left-to-right ``locally subluminal'' arrow, going instead of right-to-left, which is also valid, so long as it is in both the past and future light cones).}}
    \label{fig:Nonlocal}
\end{figure}

We can then cut out regions from the closed surface $S_1$ using the light-cones: $S_2:= S_1 \cap L(A)$  and arrive at a stronger version of Continuity of Action :
\begin{quote}
 {\bf Strong Continuity of Action (locally subluminal):} An m-model has Strong Continuity of Action {iff} local beables outside the forward and backward light-cones of a region play no role for calculating outputs in that region $P({\cal O}(A) | {\cal I}(S_2), {\cal I}(B)) = P({\cal O}(A) | {\cal I}(S_2)) ~~\forall~~ S_2$ that do not overlap with the light cones of $B$, ie $S_2 \cap L(B) = \emptyset$ (see Fig.~\ref{fig:CoA012}b). 
\end{quote}
And correspondingly,
\begin{quote}
 {\bf Weak Continuity of Action (locally superluminal):} An m-model has Weak Continuity of Action {iff} it fulfils Continuity of Action but not Strong Continuity of Action.
\end{quote}
 Weak Continuity of Action, roughly speaking, means that influences happen locally, but sometimes superluminally.  What we call Strong Continuity of Action was called Einstein locality in \cite{maudlin2012}. Local and non-local models can further be distinguished into those which are superluminal and subluminal. It it is rather uncommon to consider subluminal non-locality, but it will be helpful in what follows to clearly distinguish non-locality from superluminality. We can define subluminal non-locality by requiring that the only local beables necessary to find out what happens at $A$ are those within or on the light-cones of $A$, and adding the the light-cones of $B$ and their inside brings no further information.
\begin{quote}
 {\bf Non-locally subluminal:} An m-model is non-locally subluminal {iff} it is non-local, but local beables outside light-cones of a region play no role for calculating outputs in that region $P({\cal O}(A) | {\cal I}(L(A)), {\cal I}(L(B))) = P({\cal O}(A) | {\cal I}(L(A))$.
\end{quote}
And consequently
\begin{quote}
 {\bf Non-locally superluminal:} An m-model is non-locally superluminal {iff} it is non-local, but not non-locally subluminal.
\end{quote}
For a summary of those four terms, see Figure \ref{fig:Nonlocal}.

 We should also mention here that strong CoA is a weaker criterion than confining CoA inside the light cones, because of the requirement that one only considers regions $S_2$ which do not overlap with the light cones of region $B$. The reason for this requirement---as noted by Bell already---is that otherwise the outputs in $B$ might well provide additional information that correlates with the inputs from $S_2$ (and hence the outputs in $A$) without influences ever leaving the light-cones (see Appendix for explanation). But of course, if one extends the light cones of regions $A$ and $B$ far enough, they will always overlap. This means that one can always try to explain violations of Strong Continuity of Action by locating an origin of correlations between $A$ and $B$ in the overlap of the light cones.\footnote{This is often called a common ``cause''. However, all that is required here is a correlation, not a causation.} We will come back to this later.

\subsection{Properties of Temporal Order}
\label{ssec:TemporalOrder}

 Like with the local and temporal properties, to talk about temporal order, a model needs to have been provided with additional information. We need not only a notion of time, but also an arrow of time that allows us to distinguish past and future. This arrow of time is often not explicitly defined but implicitly assumed. Typically it comes in because we assume that an experimenter can chose a setting in the future, but not one in the past. That is, the arrow of time sneaks in with what we consider to be a possible scenario.

 Since an arrow of time is \textit{a priori} a matter of definition, we have to specify that this definition has to be consistent for all mathematically equivalent expressions of a setup. 

 There are many notions of arrows of time that have been discussed in the literature \cite{price1996time,albert2000time} and our aim here is not to unravel this discussion. We will merely note that a model needs to have such a notion for the following properties to be well-defined. 

 Like before, it is possible to have a model that just does not have a temporal order, or that does not distinguish past and future. Indeed, this is the case for many of the simplest models that we deal with, such as an undamped harmonic oscillator, or the two-body problem in Newtonian gravity. 
\begin{quote}
 {\bf Acausal:} An m-model is acausal {iff} it does not have a well-defined arrow of time, and hence no notion of past or future.
\end{quote}
 An atemporal model is also acausal---one cannot have an arrow of time without having time to begin with. This feature is exhibited in the process matrix formalism, which can even be used to model processes for which it is impossible to specify a well-defined arrow of time\cite{articleoreshkov,Oreshkov2}. One might plausibly argue that a model which isn't time-reversible necessarily has an arrow of time, and hence can't be acausal, but then we didn't say who or what defines that arrow of time, so we cannot draw this conclusion. 

 As stressed earlier, some properties of models only make sense with an arrow of time that orders times. We now come to the first of those:
\begin{quote}
    {\bf Temporally Deterministic}: A c-model is temporally deterministic {iff} it is both deterministic, and has an arrow of time, according to which calculating localisable output values at time $t'$ does not require inputs that are local beables at $t>t'$. An m-model is temporally deterministic if at least one of its c-models is.
\end{quote}
 This is a complicated way of saying that, in a temporarily deterministic model, a future state can be calculated with certainty from a past state, but not necessarily the other way round. {This notion of temporally deterministic is what is often meant by the term `deterministic'.} Note that a temporally deterministic model might have other inputs besides value-assignments for local beables. In particular, a model with all-at-once inputs may still be temporally deterministic. 
 
 We have defined temporal determinism of an m-model from the requirement that at least one of its c-models has that property, because temporal determinism is easy to remove by redefining all variables so that they mix different times, or using (partially) time-like boundary conditions. 

 In the Copenhagen Interpretation (CI), so long as no measurement occurs, the state of the wavefunction at time $t_a$ is temporally determined by the state of the wavefunction at time $t_b \neq t_a$ and the Hamiltonian operator. The state of the wavefunction after measurement, on the other hand, is generically not determined by the state of the wavefunction before measurement. Hence, the CI (c-model) is not temporally deterministic. 

 It does not follow from this that standard quantum mechanics (SQM), which we defined as [CI]$_{\rm m}$, is also not temporarily deterministic. However, this is so because---as we saw earlier---{\sc{SQM}} is not deterministic to begin with, so it cannot be temporarily deterministic either. 

 For temporal models we can further define:
\begin{quote}
    {\bf Time-reversible}: An m-model is time-reversible {iff} it is both temporally deterministic, and also deterministic under the replacement $t\to -t$, where $t$ measures time.
\end{quote}

 As mentioned earlier, this definition implicitly makes a statement about what properties we expect of time, and hence cannot stand on its own. That is not its purpose. Its purpose is rather to capture what properties physical properties like time and time-reversibility should have. 

 Time-reversibility should not be confused with invariance under time-reversal, which is a stronger requirement, but one that we will not consider here. Just because a model is time-reversible, does not mean that its time-reversed version is the same as the original.

 Next we recall a term previously-introduced in \cite{wharton2020colloquium}:
\begin{quote}
    {\bf Future Input Dependence (FID)}: A c-model has a {\sc FID} {iff}, to produce output for time $t$, it uses local beables from at least one time $t' > t$ for at least one scenario.
\end{quote}
 {\sc FID} is a property of the setup of the c-model, and may simply be a matter of choice. For example, in any time-reversible c-model, one can replace a future input with a past input and get the same outputs. We define it here because it was previously used in \cite{wharton2020colloquium} and we want to make a connection to this definition below. However, we also want to define a somewhat-stronger property:
\begin{quote}
    {\bf Future Input Requirement (FIR)}: A c-model has a {\sc FIR} {iff} there is at least one scenario for which producing outputs for time $t$ requires inputs from $t' > t$. An m-model has a {\sc FIR} if all c-models in its equivalence class have a {\sc FIR}.
\end{quote}
 Another way to phrase this is that a c-model with a {\sc FIR} has at least one scenario for which the input cannot be entirely chosen in the past.
 A c-model with a {\sc FIR} cannot be temporally deterministic. However, a model that is not temporally deterministic does not necessarily have an {\sc FIR}.

A well-known example for a Future Input Requirement is a space-time that is not globally hyperbolic, in which case the initial value problem is ill-posed. The time-evolution can then not be calculated unless further input is provided about what will happen. 

Another example might be a model which evaluates possible policy paths to limit global temperature increase to certain goals within a certain period of time (say, 2, 3, or 4\textdegree C by 2050). Such a model requires specifying the desired boundary condition in the future. Of course in this case one is dealing with hypothetical scenarios, but the example illustrates that Future Input Requirements are used in practical applications.

\subsection{Properties of Causal Order}
\label{ssec:Causal}

\begin{figure}
    \centering
    \includegraphics[width=\linewidth]{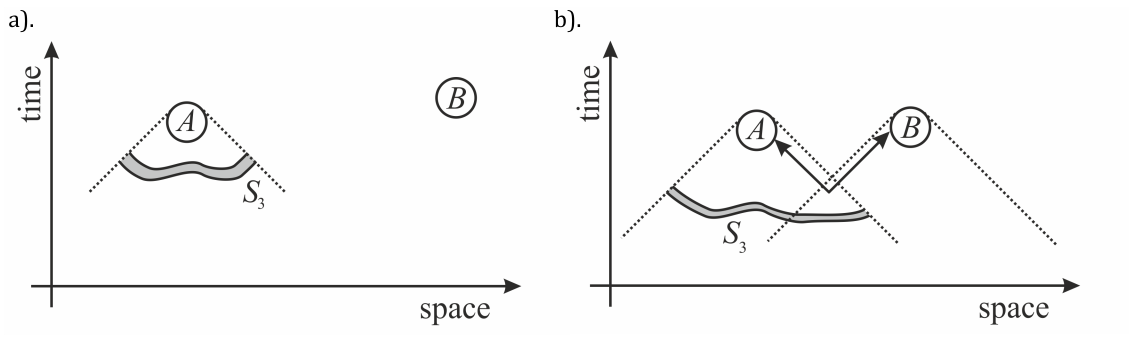}
      \vspace*{-0.5cm}
    \caption{a). Local causality. b). A subtlety of local causality.}
    \label{fig:CoA3W}
\end{figure}

 For models with a temporal order, we can distinguish the past/backward and the future/forward light-cones of a region $A$, that we will denote $L_{\rm p}(A)$ and $L_{\rm f}(A)$, respectively. It is $L(A) = L_{\rm p}(A) \cup L_{\rm f}(A)$, and we define $S_3:= S_1 \cap L_{\rm p}(A)$, the intersection of the shielding region $S_1$ with the past light-cone. With this, we can further refine Continuity of Action to a notion of local causality:
\begin{quote}
 {\bf Local Causality:} An m-model is locally causal {iff} it fulfils Strong Continuity of Action with local beables localised in the past light cone: $P({\cal O}(A) | {\cal I}(S_3), {\cal I}(B)) = P({\cal O}(A) | {\cal I}(S_3)) ~~\forall~~S_3$ that do not overlap with the light cones of $B$, ie $S_1 \cap L(B) = \emptyset$ (see Fig.~\ref{fig:CoA3W}a).
\end{quote}

 We want to stress that this notion of local causality is based on the mathematical structure of space-time, and so mixing it with other notions of causality can result in confusion. In particular, space-time causality is not \textit{a priori} the same notion of causality that is used in large parts of the philosophical literature, which concerns itself with the question of cause and effect (one currently popular realisation of which is the one based on causal models, that we will refer to as interventionist causality {\cite{pearl2009causality,glymour2001causation}}).

 According to space-time causality, of two causally related events, the one in the past is the cause by definition. One thus has to be careful when interpreting local causality using other notions of causality. This is especially important to keep in mind when we classify retrocausality.

 If one interprets ``causal'' as referring to space-time causality, then the term ``retrocausal'' suggests influences that go inside the light cones, but backwards in time. The term retrocausal, however, does not necessarily imply locality. For example, the Transactional Interpretation \cite{cramer1986transactional,kastner2013transactional}, often referred to as retrocausal, is also non-local.

We will hence proceed by endowing the classification of the four local properties, summarized in Figure \mbox{\ref{fig:Nonlocal}} with an additional temporal distinction that is either forward in time or backward in time, according to the arrow of time that we have assumed exists. We would like to stress that since we have an arrow of time, we can meaningfully distinguish forward and backward in time also outside the light-cones. The reader may want to think of the arrow of time as a preferred slicing in space-time. This will give us a total of 8 distinctions. It is then straightforward to define:
\begin{quote}
 {\bf Local Retrocausality:} An m-model is locally retrocausal {iff} it fulfils Strong Continuity of Action and has a future input requirement.
\end{quote}
\begin{quote}
 {\bf Non-local causality:} An m-model is non-locally causal {iff} it is non-local, subluminal and has no future input requirement.
\end{quote}
\begin{quote}
 {\bf Non-local Retrocausality:} An m-model is non-locally retrocausal {iff} is non-local, subluminal and has a future input requirement.
\end{quote}

As we noted already, the term ``causality'' might either refer to the light-cone structure of space-time or to interventionist causality. However, since the term ``local causality'' is extremely widely used and refers to space-time causality, we will use the term ``retrotemporal'' to refer to models with a future input requirement that do not respect the light-cone structure. This gives us the remaining four definitions:
\begin{quote}
 {\bf Locally Superluminal Retrotemporal:} An m-model is locally superluminal retrotemporal {iff} it fulfils Weak Continuity of Action and has a future input requirement.
\end{quote}
\begin{quote}
 {\bf Locally Superluminal Temporal:} An m-model is locally superluminal temporal {iff} it fulfils Weak Continuity of Action but has no future input requirement.
\end{quote}
\begin{quote}
 {\bf Non-Locally Superluminal Retrotemporal:} An m-model is non-locally superluminal retrotemporal {iff} is non-locally superluminal and has a future input requirement.
\end{quote}
\begin{quote}
 {\bf Non-Locally Superluminal Temporal:} An m-model is non-locally superluminal temporal {iff} it is non-local and superluminal, but has no future input requirement.
\end{quote}

 Since one can combine forward and backward causes to a zigzag \cite{Price2015Disentangling}, a locally retrocausal model might appear to be superluminal. However, whether such combinations are possible depends on the model.

 The above types of retrocausality and retrotemporality are properties of the mathematical structure: A future input requirement cannot be removed by a mathematical redefinition, it is therefore a property of a representation of a model. 
 An example for a locally retrocausal model might be General Relativity in space-times that have time-like closed curves. 

 However, there is a weaker notion of retrocausality that concerns the use of a c-model rather than its mathematical structure, the associated m-model. We can again distinguish four cases, which are the same as above, except that instead of a future input requirement, we merely have a future input dependence:
\begin{quote}
 {\bf Local Pseudo-Retrocausality:} A c-model is locally pseudo-retrocausal {iff} it fulfils Strong CoA, and has a future input dependence.
\end{quote}
One can similar define the terms Non-Local Pseudo-Retrocausality, Locally Superluminal Pseudo-Retrotemporarity, and Non-Locally Superluminal Pseudo-Retrotemporality, by taking the definition without the ``Pseudo'' and replacing the future input requirement with a future input dependence.

The reason we use the prefix ``pseudo'' is because according to our earlier definition a future input dependence is a matter of choice. It can be replaced with a past input, at least in principle. This does not mean that a future input dependence is unimportant, however. This is because it could be that removing the future input much increases the complexity of using the model. That is, future input dependence is linked to the question of whether a model is fine-tuned which will be further explored in the companion paper \cite{hossenfelderTK}.

 Causal and retrocausal non-locality can only be distinguished in the presence of an arrow of time, which for all practical purposes defines a space-time slicing. The same is the case for superluminal causal and superluminal retrocausal models---absent an arrow of time, they cannot be told apart, since Lorentz-transformations can convert one into the other.

 We want to stress that pseudo-retrocausality is a property of a c-model, but not a property of an m-model. 
 A retrocausal c-model cannot be temporally deterministic. A pseudo-retrocausal c-model, in contrast, can be temporally deterministic. It follows that temporal determinism is a simple way to tell pseudo-retrocausality from retrocausality. Note that pseudo-retrocausality was referred to just as retrocausality in \cite{wharton2020colloquium}. 
 The practical use of pseudo-retrocausality will be discussed in the accompanying paper \cite{hossenfelderTK}.

 The reader who at this point despairs over the many different types of retrocausality will maybe understand now why the literature on the topic is so confusing, and hopefully also why a common terminology is necessary.

\begin{figure}
    \centering
    \includegraphics[width=\linewidth]{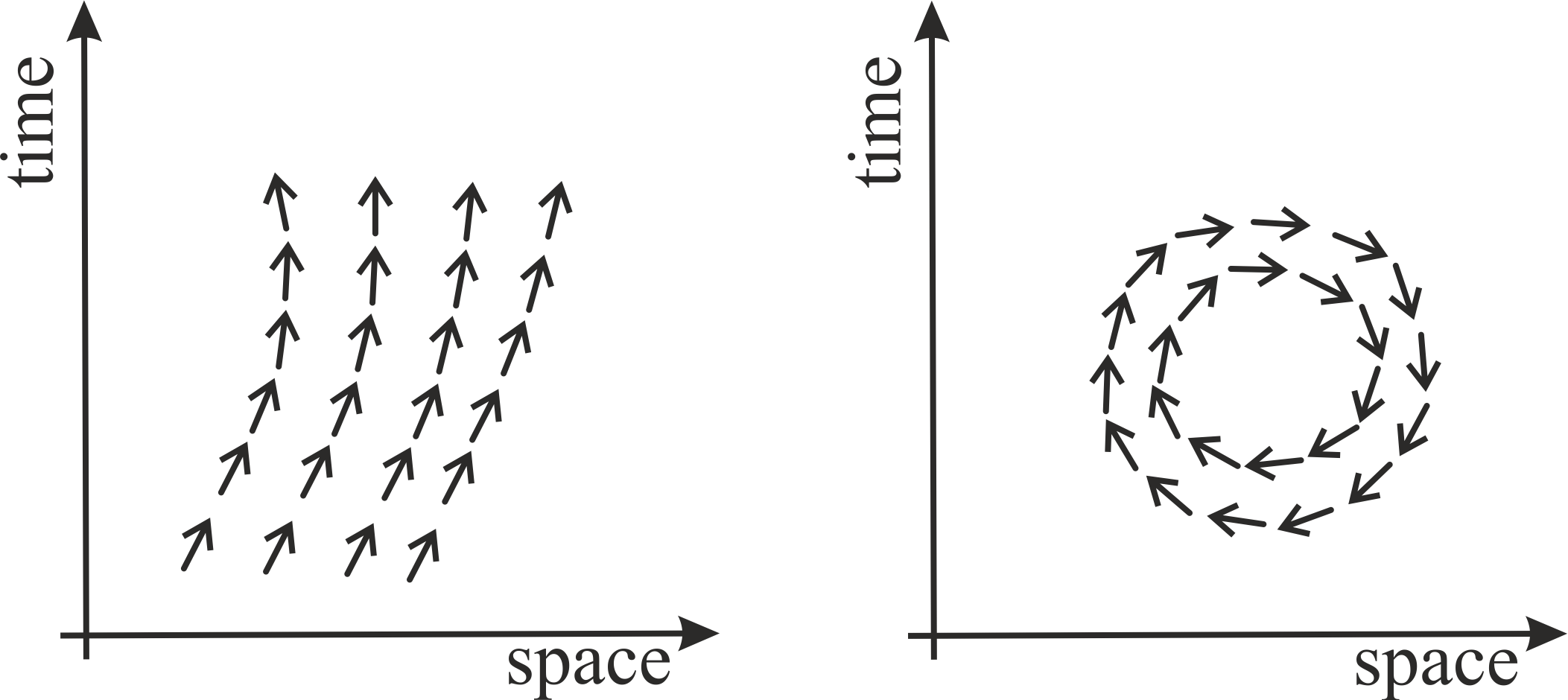} 
    \caption{Orientable and non-orientable arrows of time. {The arrows in the figure indicate the hypothetical flow of internal time of an observer, which might differ from the coordinate time. That is, the arrows are not in all places time-like, according to the coordinate time. This is supposed to illustrate the often-used example in which a spaceship that can travel faster than light in one frame actually seems to go back in time in another frame. If that was possible, one could use it to construct loops that seem to be “timelike” from the point of view inside the spaceship (i.e., permissible motion for massive objects), but not according to the coordinate time.}}
    \label{fig:aots}
\end{figure}

 Some properties about causal structure just come from the definition of the arrow of time. In particular, we have to distinguish oriented and non-oriented arrows of time. A non-oriented arrow of time is one that allows forming loops in time (Fig.~\ref{fig:aots}):
\begin{quote}
 {\bf Dynamical Retrocausality:} An m-model is dynamically retrocausal {iff} its retrocausality is due to a non-oriented arrow of time. 
\end{quote}
 Dynamical retrocausality may or may not be due to a space-time structure with closed time-like curves. It is important to emphasise that dynamical retrocausality is a property of the model, and not a property of the way the model uses inputs. Depending on how the arrow of time is defined, it may not be particularly meaningful. What makes an arrow of time meaningful, however, is not a question we want to unravel here. 
 
 Just for completeness, we also define:
\begin{quote}
 {\bf Counter-Causal:} An m-model is counter-causal {iff its time-reversed version is locally causal}.
\end{quote}

 A model with such a property would makes one strongly suspect that the arrow of time was just defined the wrong way round to begin with. However, possibly there were other reasons to define an arrow of time that way. 

 A general comment is in order here, which is that the term ``retrocausal'' is somewhat linguistically confusing. It does not so much refer to causes generally going backwards, but rather to them sometimes going against an arrow of time that was defined from something else. That is, it is really a mix of different directions of time that mark a retrocausal model, the already mentioned property that has previously been referred to as the possibility of zigzags in space-time \cite{Price2015Disentangling}. Note that the zigzag property itself is defined against the presumed-to-exist GR arrow of time.

\subsection{Agent-based Properties}
\label{ssec:Agents}

 Bell {\cite{bell1975theory}} further introduces local beables that are {\bf{controllable}}, and those which are {\bf{not controllable}}, a distinction that we will also use below. This notion is somewhat vague, but for our purposes we do not need a precise definition. We will take a controllable local beable to be one whose value can in practice be set by the action of an experimentalist---typically this is a detector setting. For a more mathematically minded definition, see {\cite{walleczek2016nonlocal}}.
 Note that for a local beable to be controllable, an experimenter need not have free will, whatever that might mean, and they also do not have to control the local beable themselves; it could be done by some kind of apparatus.

 If controllable input is correlated with observable output, we will speak of signalling. Signalling is particularly interesting if it is outside the forward light cones.
\begin{quote}
 {\bf Superluminal Signalling:} A c-model allows superluminal signalling {iff} it is superluminal and has controllable inputs which are local beables in a region $A$ that are correlated with observable outputs that are local beables in a region outside $L(A)$. An m-model allows superluminal signalling if at least one c-model in its class does.
\end{quote}

 {\sc SQM} is non-local but does not allow superluminal signalling {\cite{ghirardi1980general}}. 

 Since the time-order of space-like separated events can change under Lorentz-transformations, superluminal signalling is usually assumed to imply the possibility of signalling back in time. 
 However, non-local signalling does not necessarily imply the possibility of signalling back in time when one has a time-order given by an arrow of time. In General Relativity, for example, this may be a time-like vector field. One can then constrain non-local signals to only be forward in time relative to the vector field. For this reason, the two phenomena---signalling non-locally and signalling back in time---are distinct. 

 Let us then move on to signals that actually travel back in time: 
\begin{quote}
 {\bf Retrocausal Signalling:} A c-model allows retrocausal signalling {iff} is  retrocausal and there is at least one scenario for which observables at time $t$ are correlated with required controllable inputs from $t' > t$. An m-model allows retrocausal signalling if at least one c-model in its equivalence class does.
\end{quote}

 This retrocausality signalling could either be local or non-local. Like for the non-signalling case discussed in Section {\ref{ssec:Causal}}, causal and retrocausal non-local signalling can only be distinguished in the presence of an arrow of time, and the same is the case for temporal and retrotemporal superluminal signalling, since Lorentz-transformations can mix both cases.

 The problem with retrocausal signalling is that the observables at time $t$ are, well, observable. If they can be affected by input at a later time $t'>t$, then the result may disagree with what one had already observed. This is what opens the door to causal paradoxes.

 We will not introduce a notion of pseudo-retrocausal signalling, as that would be a technically possible definition, but rather oxymoronic. If a future input was removable and therefore not necessary for an earlier observable, then no signal was sent (though in such a case an agent might still have an illusion of signalling). 

\section{Specific Model Properties}
\label{sec:Specific}

 In this section we will now introduce properties that are specific to models typically used in the foundations of quantum mechanics.

 Using the terminology of the previous sections, we can summarise the key conundrum of standard quantum mechanics ({\sc SQM}) by saying that models mathematically equivalent to the Copenhagen Interpretation, $[{\rm CI}]_{\rm m}$, do not fulfil Strong Continuity of Action. However, {\sc SQM} also has the odd property that observables generically do not have definite values before a measurement. This opens the possibility that one may still find a physical or empirical equivalent to the Copenhagen Interpretation that fulfils Strong Continuity of Action. 

 Of course, one may be interested in interpretations or modifications of quantum mechanics for other reasons. For example, one may want to find a realist interpretation, or to return to determinism. However, the models we are here mainly concerned with are those which reinstate Strong Continuity of Action. Such models usually introduce new input variables. Using the terminology of \cite{maudlin1995three}, we will call them additional variables:
\begin{quote}
 {\bf Additional Variables:} Input values for an interpretation or modification of $[{\rm CI}]_{\rm m}$ that have no equivalent expression in $[{\rm CI}]_{\rm m}$. 
\end{quote}
 Additional variables are not necessarily localised, or even localisable, and they are not necessarily hidden, though all of that might be the case. E.g., in Bohmian Mechanics, particle positions are localisable, localised, and hidden. Bohmian Mechanics, however, as we noted previously, does not fulfil Continuity of Action. 

 One might worry here that since the variables are ``additional'', they are unnecessary to produce the same output as {\sc SQM}, and therefore just make a model's setup reducible. However, this does not have to be the case, because one normally introduces the additional variables to remove another assumption from $[{\rm CI}]_{\rm m}$. This is typically the measurement update postulate, since it is the one that leads to a violation of Strong Continuity of Action \cite{hance2022measprob}.

 The purpose of additional variables reveals itself when one interprets the violation of Strong Continuity of Action in the Copenhagen Interpretation as being due to the epistemic character of the wavefunction. One assumes that the real (ontic) state is not the wavefunction, but one that respects Continuity of Action, one just does not know which ontic state one is dealing with until a measurement was made. Quantum mechanics, in this picture, is just an incomplete description of nature. 

 We know from Bell's theorem that all such ensemble models with additional variables which determine the measurement outcome will violate an assumption to this theorem commonly known as statistical independence \cite{hance2022ensemble}:

\begin{quote}
 {\bf Statistical Independence:} An m-model fulfils statistical independence {iff} $P(\lambda|X,Y)=P(\lambda)$, where $\lambda$ is (a set of) local beables on $S_3$ and $X$ and $Y$ quantify the settings of two detectors in regions $A$ and $B$ (compare with Fig.~\ref{fig:CoA3W}a).
\end{quote}

 From this we can tell that all local interpretations or modifications of quantum mechanics can be classified by the ways in which they violate statistical independence\footnote{It is sometimes argued that the Many Worlds Interpretation is a counterexample to this claim \cite{waegell2020reformulating}. However, as laid out in the companion paper \cite{hossenfelderTK}, the Many Worlds Interpretation is either not empirically equivalent to SQM, or violates Continuity of Action.}. We will hence refer to them all as {\sc SI}-violating models. 

 Statistical independence is also sometimes referred to as ``measurement independence,'' or the ``free choice assumption'' or the ``free will assumption'' in Bell's theorem. In rare occasions we have seen it being referred to as ``no conspiracy''. In recent years, theories which violate statistical independence have also been dubbed ``contextual'' \cite{Kupczynski:2022lls}, though the class of contextual models is larger than just those which violate statistical independence (there is more ``context'' to an experiment than its measurement setting). 

 Not all models that are being used in quantum foundations reproduce all predictions of quantum mechanics. Many of them only produce output for certain experimental situations, typically Bell-type tests, interferometers, or Stern-Gerlach devices. We can then ask, in a hopefully obvious generalization of the above classification, whether these models are either representations or interpretations of $[{\rm CI}]_{\rm m}$ as it applies to the same experiments.

 Traditionally, a distinction has been made between {\sc SI}-violating models that are either retrocausal or superdeterministic. But this distinction has remained ambiguous for three reasons. 

 First because---as we have seen already---retrocausality itself has been used for a bewildering variety of cases. If one then defines superdeterminism as those {\sc SI}-violating models which are not retrocausal, one gets an equally bewildering variety.
 Second, since Bell (who coined the term ``superdeterminism'') did not distinguish retrocausality from superdeterminism, one could reasonably argue that superdeterminism should be equated with {\sc SI}-violation in general, and then consider retrocausality to be a variant of superdeterminism.
 Third, not everyone agrees that superdeterministic models have to be deterministic to begin with \cite{Sen2020Superdet1,Sen2020Superdet2}.

 There is no way to define the term so that it agrees with all the ways it has previously been used. We will therefore just propose a definition that we believe agrees with the way it has most widely been used, based on the following reasoning:
 
\begin{enumerate}
    \item We are not aware of any superdeterministic model which is not also deterministic. Leaving aside that it is terrible nomenclature to speak of ``non-deterministic superdeterminism'', there is not even an example for it. For this reason, we will assume that superdeterministic models are deterministic.
    \item We want a model that fulfils Strong Continuity of Action, because this is the major reason why violations of statistical independence are interesting, and it is also the context in which Bell coined the expression.
    \item Most of the literature seems to consider superdeterminism and retrocausality as two disjoint cases, so we will do the same, even though Bell seems to not have used this distinction when he coined the term. This point together, with the previous one, implies that the model has to be locally causal.
    \item We want a distinction that refers to the m-model, not to its particular realization as a c-model to avoid new ambiguities.
    \item The model should reproduce the predictions of quantum mechanics, at least to a reasonable extent. This assumption is relevant because without it Newtonian mechanics would also be superdeterministic which makes no sense.
    \item It's a one-world model that violates statistical independence. 
\end{enumerate}
Some readers might argue that requirement 6 is strictly speaking unnecessary because it follows from Bell's theorem given the previous 5 requirements. However, since we did not explicitly list the assumptions to Bell's theorem, and it is somewhat controversial which of those have to be fulfilled in any case, we add it as an extra requirement.

Taking this together, we arrive at the following definition:
\begin{quote}
 {\bf Superdeterminism:} An m-model with additional variables which is deterministic, locally causal, violates statistical independence, and is empirically equivalent to standard quantum mechanics (is in [{\sc{CI}}]$_{\rm e}$).
\end{quote}

 With this definition, a superdeterministic model is distinct from a retrocausal model. However, a superdeterministic c-model can still be pseudo-retrocausal. This distinction has previously been made in \cite{nikolaev2022aspects}. The authors of this paper used the term ``Soft Superdeterminism'' for what in our nomenclature would be a pseudo-retrocausal superdeterministic c-model, and they use the term ``Hard Determinism'' which in our nomenclature would be a not pseudo-retrocausal, superdeterministic c-model.
 In both cases though, the m-class belonging to the model is superdeterministic, and not retrocausal. 
\begin{figure}
    \centering
    \includegraphics[width=\linewidth]{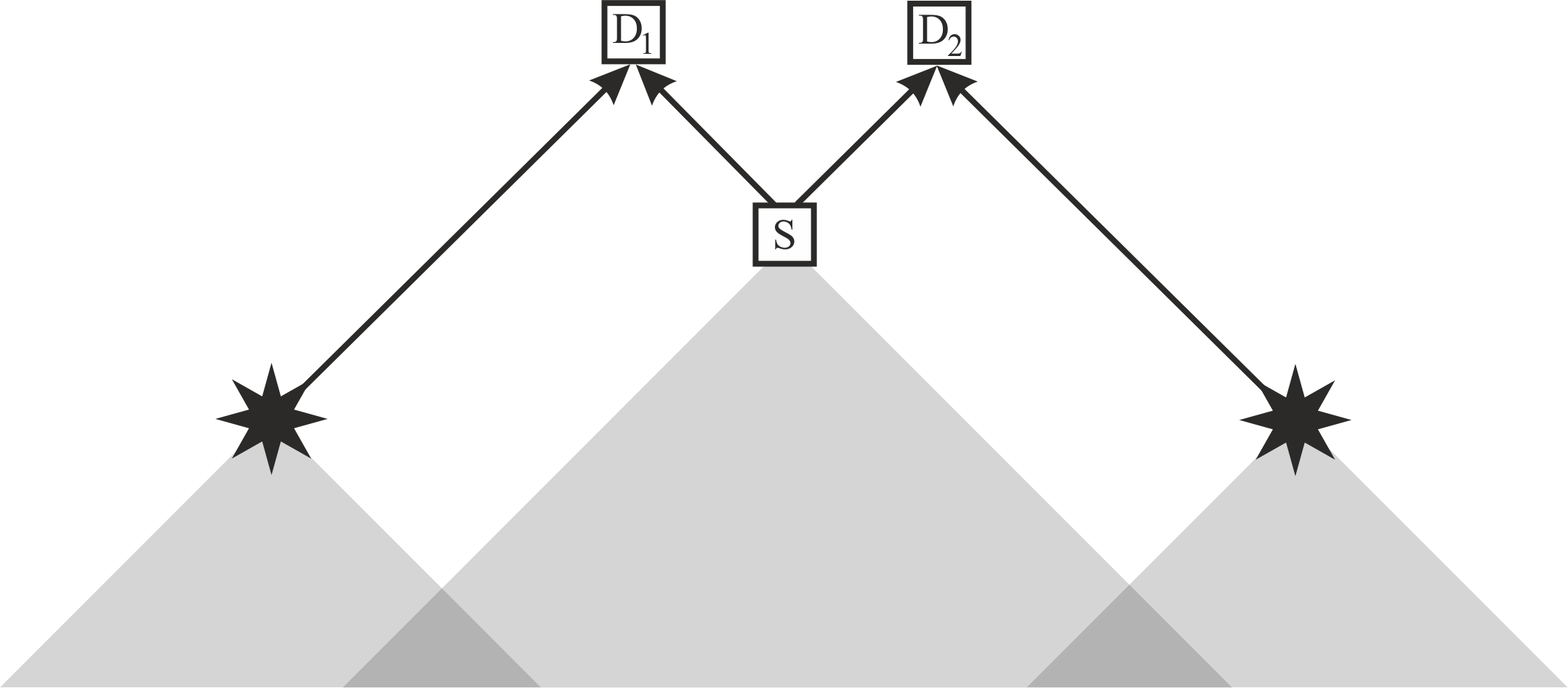} 
    \caption{Sketch of Cosmic Bell test with source S and two detectors. The settings of the detectors D$_1$ and D$_2$ are chosen by using photons from distant astrophysical sources, usually quasars. Any past cause that could have given rise to the observed correlations must then have been very far back in time. Light cones indicated by gray shading.}
    \label{fig:cbt}
\end{figure}

 The notion of common-cause superdeterminism refers even more specifically to a certain type of superdeterministic models that are not pseudo-retrocausal. In these models, one assumes that the additional variables which determine the measurement outcomes in a Bell-type experiment are correlated with the settings, because they had a common cause in the past. 

We want to emphasise that the additional requirement of common-cause superdeterminism is that there was a \mbox{\emph{common}} cause, a more or less localised event that serves as what some would call an ``explanation'' for the correlation that gives rise to violations of statistical independence. This is opposed to the general superdeterministic models in which the correlation does not require explanation, but rather \mbox{\emph{is}} the explanation (for our observations). The correlation that gives rise to violations of statistical independence of course always has a space-time cause---in the most extreme case that would be the initial condition at the beginning of the universe. But in general the hidden variables and detector settings have no common cause in the interventionist sense, nor do they need have to have one.

That said, this particular type of common-cause superdeterminism can be tested by using methods to choose the detector settings that are so far apart from each other that any common cause would have had to be in the very early universe. This is the idea behind the Cosmic Bell test \cite{gallicchio2014testing,rauch2018cosmicquasar}, sketched in Fig.~\ref{fig:cbt}.

 However, while common-cause superdeterminism is a logical possibility, we are not aware of any superdeterministic model which is of this type, or of anyone who has advocated such a model. 

 Unfortunately, the results of Cosmic Bell tests are often overstated in the literature, quite frequently claiming to ``close the superdeterminism loophole'', rather than just testing a particular type of superdeterministic model, which no one works on to begin with. 

 The relation between pseudo-retrocausality and finetuning (a.k.a. ``conspiracies'') will be explored in a companion paper \cite{hossenfelderTK}.

\section{Classification}
\label{sec:Class}

 We want our classification scheme to be practically useful, so we will here provide a short guide to how it works. 
 
 The question we want to address is this:
 
 Supposed you have a c-model for quantum mechanics---that is the thing you are doing your calculations with---what should you call it? We will assume that your model is presently empirically equivalent to standard quantum mechanics {in the non-relativistic limit}. (If it isn't, you have bigger problems than finding a name for it.) You then proceed as follows:
\begin{enumerate}
    \item To classify the model, you first have to make sure that its setup irreducible. If it is reducible, the setup can't be classified, so please remove all assumptions that are not necessary to calculate outputs. Assumptions that state the ``physical existence'' (whatever that may be) of one thing or another are typically unnecessary for any calculation. 

    \item Figure out whether your model is physically equivalent to {\sc SQM}. If it is not, it's a {\bf modification}. If it is, it's an {\bf interpretation}.

\item Figure out whether your model is mathematically equivalent to any already known interpretation. If it is, your model is a {\bf representation} (of the interpretation it is mathematically equivalent to). If it's not, you have a representation of a new interpretation.

    \item  Go through the list of properties in Sections \ref{sec:Prop} and \ref{sec:Specific}, and note down whether your model has them, starting with the m-model properties, then the c-model properties. 
\end{enumerate}

\section{Summary}
\label{sec:Sum}
 
 We have proposed a classification scheme for models in the foundations of quantum mechanics. Its most central element is the distinction between different types of models: calculational, mathematical, physical, and empirical. After distinguishing these different classes of models, we have defined some of their properties that are discussed most commonly in the foundations of quantum mechanics, with special attention to those concerning locality and causality. We hope that the here proposed terminology can aid to clarify which problems of quantum mechanics can and cannot be solved by interpretation. 
 
\textit{Acknowledgements:} We want to thank Louis Vervoort and Ken Wharton for valuable feedback and all participants of the 2022 Bonn workshop on Superdeterminism and Retrocausality for helpful discussion.
JRH acknowledges support from Hiroshima University's Phoenix Postdoctoral Fellowship for Research, the University of York's EPSRC DTP grant EP/R513386/1, and the UK Quantum Communications Hub, funded by the EPSRC grants EP/M013472/1 and EP/T001011/1.

\subsection*{Acronyms}

\begin{tabular}{cc}
  {\sc CI}: & Copenhagen Interpretation \\
  {CoA}: & Continuity of Action\\
{\sc FID}: & Future Input Dependence\\
{\sc FIR}: & Future Input Requirement\\
{\sc SQM}: & Standard Quantum Mechanics\\
{\sc SI}: & Statistical Independence
\end{tabular}

\subsection*{Appendix}

If the region $S_3$ was allowed to intersect with the inside of the past-lightcone of $B$, then in a theory which is not temporally deterministic, correlations could be created later which were not contained on $S_3$. In this case then, local beables at $B$ could provide extra information for what happens at $A$ though the information got there locally and inside the light-cones. For illustration, see Figure \ref{fig:CoA3W}b. 

\bibliographystyle{unsrturl}
\bibliography{ref.bib}

\end{document}